\documentstyle [12pt] {article}

\begin{document}
\title {\large \bf THE QUANTUM DE RHAM COMPLEXES \\ ASSOCIATED WITH
  $ SL_h(2) $ }
\author {VAHID. KARIMIPOUR}
\date { }
\maketitle
\begin {center}
{\it  Department of Physics , Sharif University of Technology\\
P.O.Box 11365-9161 Tehran, Iran\\
Institute for studies in Theoretical Physics and Mathematics
\\ P.O.Box 19395-1795 Tehran, Iran \\
Email. Vahidka@irearn.bitnet}.
\end {center}
\vspace {1cm}
\begin {abstract}
{Quantum de Rham complexes on the quantum plane and the quantum group
itself are  constructed for the Zakrewski deformation of $ Fun ( SL(2)) $.
A new deformation of the two dimensional Heisenberg algebra is also
 obtained
which can be used to construct models of h-deformed quantum mechanics.}
\end{abstract}
\noindent
\newpage
{\large \bf I. Introduction}\\

There are two quantum group structures which admit a central determinant
 on the space of 2 by 2 matrices. The familiar $ GL_q(2) $ [1] and
the recently
discovered $ GL_h(2) $ [2] .
The later was found by Zakrewski by applying the
general construction of the Leningrad school [3] to the
following R matrix:
\begin{equation} R = \left(\begin {array}{llll} 1 & h & -h & h^2  \\
0 & 1 & 0
& h \\ 0 & 0 & 1 & -h \\ 0& 0 & 0 &1  \end {array}\right) \end {equation}
and a new deformation of the algebra of polynomial functions on $ GL(2)$
was constructed. This algebra which we denote $ A_h$ is generated
by 1 and the elements of a quantum matrix
$ T = \left(\begin{array}{ll} a & b \\ c & d  \end{array}\right) $
 ,subject to
the following relations:
$$ [ c , a ] = h c^2 \hskip 2cm [ b , a ] = h( D - a^2 ) $$
\begin{equation} [ c , d ] = h c^2 \hskip 2cm [ b , d ] = h( D - d^2
 ) \end {equation}
$$ [ a , d ] = h(a - d) c  \hskip 2cm [ c , b ] = h( ac + cd ) $$
Where $ D= ad-bc-hac $ is called the quantum determinant of T .
The classical limit is obtained by setting $ h $ equal to zero.
$ A_h $ is turned into a Hopf algebra by the mappings:
$$ \Delta \left( \begin{array}{ll} a & b \\ c & d  \end{array}
\right) =  \left( \begin{array}
{ll}a\otimes a + b\otimes c &a\otimes b + b\otimes d
\\ c\otimes a + d\otimes
c & c \otimes b +d\otimes d \end{array}
\right) $$
\begin {equation}\epsilon \left( \begin{array}{ll} a & b \\ c & d
\end{array}
\right) = \left( \begin{array}{ll} 1 & 0 \\ 0 & 1  \end{array} \right)
\end{equation}$$ S \left( \begin{array}{ll} a & b \\ c & d  \end{array}
\right)
= \left( \begin{array}{ll} {d - h c}  & {-b + ha - hd + h^2 c}  \\ {-c} &
{a+hc}   \end{array} \right) $$
One can check that $ D $ is central and hence can be set equal to unity (
$ ad - bc -hac = 1 $ ) , $ A_h $ is then a deformation of Fun(SL(2)).
( For generic D , $ A_h $ is a deformation of $ GL_2 $ ). The quantum
determinant is also
grouplike which means that $ \Delta D = D \otimes D $  and $ D(TT') =
D(T)D(T') $ for two matrices $ T $ and $ T' $  the
elements of which seperately satisfy (2)  while  $ [ t_{ij} ,
 t'_{kl} ] = 0 $.
The algebra dual to $ A_h $ has also
been constructed in [4] and a new deformation of $ U(sl(2)) $  has been
obtained . $ U_h (sl(2)) $ is generated by three elements $ H , X ,
$ and $ Y
$ satisfying :
$$ [ H , X ] = { 2 sinh\ \  hX \over h } $$
$$ [ H , Y ] = -Y(Cosh \ \ hX) - (Cosh\ \  hX)Y $$
$$ [ X , Y ] = H $$
For the Hopf structure see ref. [4].

The noncommutative differential calculus associated with $GL_q(2) $ is well
known ([5]) ,i.e: the following things are known:

a) - The pair of spaces denoted by $R_q(2)$ and $ R^*_q(2)$ on which $
GL_q(2) $ acts as a quantum automorphism group.

b) - A consistent differential calculus on $R_q(2)$ which makes $ R_q(2) $
and $ R^*_q(2) $ as parts
of a graded differential algebra, namely the quantum de Rham complex $
\Omega(R_q(2)$.

c) - The quantum de Rham complex of $ SL_q(2) $, denoted by $
\Omega(SL_q(2))$.

d) - The deformation of the Heisenberg algebra of coordinates and momenta
which
is the result of suitable identification of momenta with the derivatives $
\partial_x $ and $\partial_y$ of part b) . We call this algebra
$ H_q( x,y ;p_x
,p_y )$.

In this letter we construct the corresponding objects for the
 quantum matrix
group $ SL_h(2) $ , which we denote respectively by $ R_h(2)
, R^*_h(2) , \Omega(R_h(2)) , \Omega_(SL_h(2) ) $ and $ H_h ( x,y ; p_x
,p_y ) $.
On the mathematical side, this letter provides a new simple example
for the study of non-commutative differential geometry.
On the physical side
Our construction,especially the new deformation of the 2 dimensional
Heisenberg
algebra can be used to construct new models of h-deformed quantum mechanics.

The structure of this paper is as follows : We construct $ \Omega [R_h(2)]$
 in
section II , and $ H_h (x,y ;p_x,p_y ) $
in section III.
Building the quantum de Rham complex $ \Omega [ R_h(2) ] $ , we
then extend a la Manin [5] the differential graded algebra homomorphism
$$ \Omega [ R_h(2) ] \longrightarrow F [ SL_h(2) ] \otimes \Omega
[ R_h(2) ]  $$
( where$ F [ SL_h(2) ] $ denotes the ring of polynomial functions on
 $SL_h(2)$ )
to
$$ \Omega [ R_h(2) ] \longrightarrow \Omega [ SL_h(2) ] \otimes
\Omega [ R_h(2) ]  $$
$\Omega [ SL_h(2) ]$ then defines the quantum de Rahm complex of
 $ SL_h(2) $ .

{\large \bf II. The Quantum De Rham Complex $ \Omega [ R_h(2) ] $}\\

First we note that $ GL_h(2) $ can be understood as a " quantum
automorphism
group " [5] of a new pair of " noncommutative linear spaces ",
$ R_h ( 2 ) $  and $ R_h(2)^* $  defined as follows:
\ $ R_h(2)  $\ \ is the two dimensional quantum plane with coordinates
 $ x $ and $ y
$ with the relation:
\begin{equation} xy-yx = -h y^2 \end {equation}
$ R_h(2)^*$\ \  is the dual quantum plane with coordinates
$ \xi $ and $ \eta $
and relations : \begin{equation} \xi ^2 = -h \xi \eta \hskip 2cm
\eta ^2 = 0 \hskip
2cm \xi \eta + \eta \xi = 0  \end{equation}
Note the peculiarity of $ \xi^2 $ which is not equal to zero , as in the
usual quantum planes associated with $ GL_q(2) $.
It is then seen that the coactions ( change of coordinates ) :
\begin{equation}   \left (\begin{array}{c} {x'} \\ {y'}\end{array}\right)=
\left( \begin{array}{ll} a & b \\ c & d \end{array} \right)
\otimes \left(\begin{array}{c} x \\ y \end{array} \right ) \hspace {1.5 cm}
\left (\begin{array}{c} {\xi '} \\ {\eta ' }\end{array}\right ) = \left(
\begin{array}{ll} a & b \\ c & d \end{array} \right)
\otimes \left(\begin{array}{c} \xi \\ \eta \end{array} \right ) \end
{equation}
preserve (4,5) provided that relations (2) hold.
All the properties of $ SL_h(2) $ acquire now a natural setting . In
particular the group property  expresses the composition of two
 automorphisms  and D can be
calculated from the formula $ \xi '\eta ' = D \otimes \xi \eta $,
exactly as
in the classical definition of the determinant via a volume form .
In the next section we construct a covariant diffrential calculus
[5,6] on $
R_h
(2) $ interpreting $ \xi $ and $ \eta $ as the noncommutative analogues
of $ dx
$ and $ dy $ respectively.
We begin from scratch relying mainly on covariance.
As we will see, contrary to the standard $ SL_q(2) $, there exists a
unique
differential calculus on $ SL_h(2) $.
\vskip 1cm
{\bf A. The relations between coordinates and differentials.}

We assume that each of the monomials $ x_i \xi_j \ \  ( x_1 = x ,
\  x_2 = y ,\  \xi_1\  =
\xi \ , \xi_2\  = \eta $\  )  can be expressed as a linear combination
of $ \xi_j x_i $ \ '  s . i.e:
\begin{equation}\left( \begin{array}{c} x \xi \\ x \eta \\ y\xi \\ y\eta
\end{array} \right) = \left( \begin{array}{llll} c_{11} & c_{12} & c_{13}&
c_{14}\\ c_{21}&c_{22}&c_{23}&
c_{24} \\ c_{31}&c_{32}&c_{33}&c_{34} \\ c_{41}&c_{42}&c_{43}&c_{44} \end
{array}
\right )  \left( \begin{array}{c} \xi x \\ \eta x \\ \xi y \\ \eta y
\end{array} \right) \end{equation}
where the entries of the above matrix which we denote by C are complex
numbers.
We now demand that:

1) Taking differentials from both sides of (7) reproduces the relations
(5) and

2) The relations (7) are preserved under change of the coordinates (6)

The above two requirements specify the matrix C modulo a parameter:
$$ C = \left( \begin{array}{llll}-1-2\beta  & \beta h  & -\beta h  &
-\beta h^2 \\  0  & -\beta & -1-\beta &
\beta h  \\ 0 & -1-\beta & -\beta & -\beta h  \\
0 & 0 & 0 & -1-2\beta  \end{array}
\right ) $$
The parameter $ \beta $ can be determined by checking relations among cubic
monomials.
In particular one should resolve the overlap ambiguity:From (7) we see that:
$$ x(\eta \xi ) = - \beta( 1+2\beta) \xi \eta x + h ( 1+\beta)^2
 \xi \eta y$$
and
$$ (x\eta) \xi  = ( - \beta( 1+2\beta) + ( 1+\beta )^2 ) \xi \eta x $$
which uniquely constrains $ \beta $ to be equal to -1.
The final result is :
$$ x \xi = \xi x - h ( \eta x - \xi y ) + h^2 \eta y $$
$$ x \eta = \eta x - h \eta y $$
\begin{equation} y \xi = \xi y + h \eta y  \end{equation}
$$ y \eta = \eta y $$
In this way $R_h(2)$ and $R_h(2)^*$ can be considered as parts of a
 differential
graded algebra $ \Omega [ R_h(2) ] $ , the quantum de Raham complex of
the quantum plane
$ R_h(2) $ . Note that the uniqness of $ \beta $ makes $ \Omega [R_h(2)
] $ and consequently
$ \Omega [SL_h(2)] $ unique , in sharp contrast with the case of
 $ SL_q(2) $
\vskip 1cm
{\bf B. The Derivatives }

Beore proceeding we would like to discuss the meaning of covariance in
noncommutative differential calculus of Wess and Zumino [6].
As regards the covariance or bicovariance in differential calculus on
 quantum
groups, there is a rather abstract formalism developed by Woronowics
[7] which

generalizes the ordinary bicovariance of the calculus on Lie groups under
left and right action of the group on itself.
However as regards the
differential
calculus on the quantum space, the language of Wess and Zumino [6]
is more accesible to physicists.The ultimate goal of this approch is to
build a model of quantum field theory on a noncommutative spacetime the
symmetry of which is the 4-dimmensional deformed poincare group.
As I understand covariance here means that all the relations between
coordinates $ x_i$ , differentials $ dx_i$ and  derivatives $\partial_{x_i}
$ etc. must preserve their form when one changes the coordinates by
$x_i \longrightarrow x'_i = T_{ij} x_j $, where the matrix T is an
element of the quantum group acting on the quantum space.
If we accept such a meaning of covariance then we face the difficulty
that the very definition of a partial derivqative acting from the
left is not covariant. This is a completely general problem and is due to
the fact that for a
quantum matrix $ T $ the relation $ (T^{-1})^t =  (T^{t})^{-1} $where$ T^t$
is the transpose of $ T $, is not valid .
In appendix A ) We explain in detail this problem for the simplest case ,i.e:
the quantum space associated with $ GL_q(2) $.
As explained in appendix A) covariance can be maintained if one defines all
the partial derivatives and the differentils to act from the right.We adopt
this strategy in the present letter.
Therefore in what follows the derivatives $\partial_x $ and $ \partial_y $
always act from the right .
We assume $ x \stackrel \leftarrow {\partial_x } =  y \stackrel \leftarrow
{\partial_y} = 1 ,
y\stackrel \leftarrow {\partial_x} = x \stackrel \leftarrow {\partial_y}
= 0 $
and the quantum Liebnitz rules  [6] : $$ (fg)\stackrel \leftarrow {d} = f
(g\stackrel \leftarrow {d}) + ( f\stackrel \leftarrow {d} ) g $$
\begin{equation} (fg) \stackrel \leftarrow {\partial_i} = f (g\stackrel
\leftarrow {\partial _i}) + ( f\stackrel \leftarrow {\partial_k} )
(g)\stackrel
\leftarrow {O_i^k} \end{equation}
where $\stackrel \leftarrow {O_i^k} $ are linear operators acting from the
right on functions of $ x $
and $ y $ . The diffrential $\stackrel \leftarrow {d} $ is $ \stackrel
\leftarrow {d} = \stackrel \leftarrow {\partial_x} \xi + \stackrel
\leftarrow
{\partial _y} \eta $.
All the relations between $ x , y , \partial_x $ and $ \partial_y $ that
will be derived are covariant under the
change of coordinates (6) .
Hereafter for notational convenience we drop the arrows from above the
symbols .
For the basic variables $ x $ and $ y $ let us parametrize the action of
the operators \ $ O_i^k $ \ as
follows :
$$  (x) O_1^1 = \alpha_1 x + \beta_1 y \hskip 2cm  (y) O_1^1 =
\mu_1 x + \nu_1
y $$
$$  (x) O_1^2  = \alpha_2 x + \beta_2 y \hskip 2cm (y) O_1^2 = \mu_2 x +
\nu_2 y $$
$$  (x) O_2^1  = \alpha_3 x + \beta_3 y \hskip 2cm (y) O_2^1  =
\mu_3 x + \nu_3
y $$
$$  (x) O_2^2 = \alpha_4 x + \beta_4 y \hskip 2cm  (y) O_2^2 =
\mu_4 x + \nu_4
y $$
where the parameters are to be determined.

Acting on $ xy $ by $ d $  from the right and using
(8) we obtain : \begin{equation} ( xy ) d = x \eta + \xi y = x\eta + y \xi -
 h y \eta \end{equation}
On the other hand
 \begin{equation} ( xy ) d = ( xy ) ( \partial_x \xi + \partial
_y \eta ) = (
\mu_1 x +
\nu_1 y ) \xi + ( x + \mu_3 x + \nu_3 y ) \eta \end{equation}
Comparing (10)
and (11) we find : $$ \mu_1 = \mu _3 = 0 \hskip 1cm \nu_1 = 1
\hskip 1cm \nu _3 = - h $$
We next repeat the above calculation with $ xy $  replaced by $ yx $,$ x^2
$  and $ y ^2 $ . The results are respectively :
$$ \alpha_2 = \beta_2 = 0 \hskip 1cm \alpha_4 = 1 \hskip 1cm \beta_4 = h $$
$$ \mu_2 = \nu_2 = \mu_4= 0 \hskip 1cm \nu_4 = 1  $$
$$ \alpha_1 =  1 \hskip 1cm \beta_1 = - \alpha_3 = h \hskip 1cm
\beta_3 = h^2 $$
Therefore we will have:
$$ ( x )O^1_1 = x - hy \hskip 2cm( y ) O^1_1 = y $$
$$ ( x )O^2_1 = 0 \hskip 2cm     ( y ) O^2_1 = 0 $$
 \begin{equation}  ( x )O^1_2 = hx + h^2 y \hskip 2cm( y ) O^1_2 =
-h y \end{equation}
$$  ( x )O^2_2 = x + hy \hskip 2cm    ( y ) O^2_2 = y $$
Combination of these relations with the Leibnitz rule (9) leads to th
e following formulas for the
derivatives of quadratic monomials:
$$ (x^2 ) \partial_x = 2x -hy \hskip 2cm (x^2 ) \partial_y = h x + h^2 y $$
$$ (xy ) \partial_x = y \hskip 2cm (xy ) \partial_y =  x - h y $$
 \begin{equation} (yx ) \partial_x = y \hskip 2cm (yx ) \partial_y =
x + h y \end{equation}
$$ (y^2 ) \partial_x = 0 \hskip 2cm (y^2 ) \partial_y = 2y $$
One can also check two facts : First the compatibility of (13)
with the defining
relations of $ R_h (2) $ ( i.e: $ r \partial_x = r \partial_y = 0
$ where  $ r = xy - yx + hy^2 = 0 $ )
and second the covariance of (13) under change of the coordinates (6).
\vskip 1cm
{\bf C. Relations between the derivatives}

Comparision of the actions of $ \partial _x \partial _y $ and
 $ \partial_y
\partial _x $ on the monomials $ x^2 , xy $  and $ y^2 $ yields :
 \begin{equation} [\partial_x , \partial_y ] = -h\partial_x^2\end{equation}
\vskip 1cm
{\bf D. Relations between the coordinates and the derivatives}

Using (9 and 12) one can verify the following identities where $ f $
denotes an
arbitrary function on $ R_h (2) $.
$$ (fx)\partial_x = f + (f\partial_x)(x-hy) $$
$$ (fx)\partial_y = (f\partial_x)(hx + h^2 y) + ( f\partial_y)(x+hy) $$
 \begin{equation} (fy)\partial_x = (f\partial_x)y  \end{equation}
$$ (fy)\partial_y = f - h(f\partial_x)y + (f\partial_y)y $$
which leads to:
$$ [ x , \partial _x ] = 1 - h \partial _x y  $$
$$ [ x , \partial _y ] = h ( \partial _x x + h \partial _x y +
\partial _y y ) $$
 \begin{equation} [ y , \partial _x ] = 0 \end{equation}
$$ [ y , \partial _y ] = 1 - h ( \partial _x ) y $$
We add the relations (4) and (14) to this set
and denote the algebra generated by the symbols $ x , y , \partial_x $ and
$ \partial_y $ by  $B_h $.
\vskip 1 cm
{\large \bf III) A  New  Deformation of the 2  Dimensional
Heisenberg Algebra }

As it stands the algebra $ B_h $  can not be interpreted as a deformation
of the 2 dimensional Heisenberg Algebra by simply identifying
$ \partial_x $ and $ \partial_y $ with the momenta $ip_x $ and $ ip_y$
since the hermiticity of the coordinates and the momenta is not compatible
with the relations (16) . ( In fact the only obstruction is the second
equation of (16) ).
However for  pure imaginary h  the algebra $ B_h $ admits the following
involution ( conjucation ) :
$$ x^{ \dagger } = x + hy \hskip 2cm y^{\dagger }= y $$
 \begin{equation} \partial_x^{\dagger} = - \partial_x  \hskip 2cm \pa
rtial_y^{\dagger} = - (
\partial_y + h \partial_x )  \end{equation}
which allows us to define the hermitian operators
$$ \hat x = x + {h\over 2} y
\hskip 2cm \hat y = y $$
 \begin{equation} \hat p_x = i \partial_x \hskip 2cm \hat p_y =
i( \partial_y + {h\over 2 }
\partial_x ) \end{equation}
It is then seen  ( after a little rearrangement ) that the following
 relations
hold, where we have set $ h=ih' , h'\in R $ .
$$ [ \hat x , \hat y ] = -ih' \hat y^2 $$
$$ [ \hat p_x , \hat p_y ] = -ih'\hat p_x^2 $$
$$ [ \hat x , \hat p_x ] = i( 1 - h' \hat p_x \hat y )$$
\begin{equation} [ \hat y , \hat p_y ] = i( 1 - h' \hat p_x \hat y )
\end{equation} $$ [ \hat y , \hat p_x ] = 0 $$
$$ [ \hat x , \hat p_y ] = {ih'\over 2 }(\hat p_x \hat x + \hat x \hat p_x+
\hat p_y \hat y + \hat y \hat p_y )  $$
One can verify directly the Jacobi Identity and the compatibility
of (19) with hermiticity of $ \hat x , \hat y , \hat p_x $ and $ \hat p_y $

The classical and quantum dynamics of a particle moving on a quantum line
has been studied by Aref'eva and Volovich [8] and Schewnk and Wess [9] . It
may
be interesting to extend their studies  to 2 dimensions using (19) as
the deformed
algebra of observalbes.
\vspace {1 cm}

{\large \bf IV - De Rham Complex of $ SL_h(2) $}

The quantum de Rahm complex of $ SL_h(2) $ , $ \Omega [ SL_h(2) ] $ is
generated by the elements $ a, b, c, d $ and
their differentials $ \alpha , \beta , \gamma , \delta $ respectivly.
 The relations between these elements can be
determined by regarding $ \Omega [ SL_h(2) ] $ as an
 automorphism of $ \Omega [ R_h(2) ] $.
 \begin{equation} \Omega [ R_h (2) ] \longrightarrow \Omega [ SL_h(2) ]
 \otimes \Omega
 [ R_h ( 2 ) ] \end{equation}
However this approach which has been explained by Manin [5]
can not determine  directly all the cross commutation relations betwe
en the elements $ a , b ... $
and their differentials. As explained in [5] one can postulate linear
relations between $ a_i da_j  $ and $  (da_j) a_i  $ ( where $ a_i =
a, b ... $ ) and constrain the coefficents by the same
technique that was used to determine the de Rahm complex of $ R_h(2)
 $. (see eqs. (7)
and the discussion following it ). For $ SL_h(2) $ however the required
 computations are lenghty and tedius.

We follow a much simpler  and yet equivalent approach inspired
by the work of
ref. [10] . That is we postulate the cross commutation
relations from the outset and then take their differentials to determine
the relations
between the one forms themselves.
Let us postulate in a by now familiar notation the following relations
 \begin{equation} R_{12} dT_1 T_2 = T_2 dT_1 R_{12}\end{equation}
The consistency of this relations with the defining relations (2)
hinges heavily
on a nice property of the R matrix (1),namely:
 \begin{equation} R_{12}R_{21} = R_{21}R_{12} = 1 \end{equation}
where $ R_{21} = PR_{12}P $ \ \  and  $ P $ is the permutation matrix.
Using this property we find from (21) that
 \begin{equation} R_{12}^{-1} dT_2 T_1 = T_1 dT_2 R_{12}^{-1}\end{equation}
Combining (21) and (23) it is easy to check that
 \begin{equation} d ( R_{12}T_1 T_2 - T_2 T_1 R_{12} ) = 0 \end{equation}
Upon taking the differential from both sides of (21) we will have:
 \begin{equation} R_{12} dT_1 dT_2 + dT_2 dT_1 R_{12} = 0 \end{equation}
Solution of (21) yields the following cross commutation relations:
$$ a \alpha = ( \alpha - h \gamma) a + h ( \alpha+h \gamma) c \hskip 1cm
b\alpha = ( \alpha - h \gamma ) ( b-ha) + h ( \alpha+h\gamma) ( d-hc ) $$
$$ c \alpha = ( \alpha + h \gamma) c \hskip 3cm
d\alpha = ( \alpha + h \gamma ) ( d-hc) $$\\
$$ a \gamma =  \gamma ( a - hc ) \hskip 1cm
b\gamma  =  \gamma (b - hd -h (a-hc)) $$
$$ c \gamma =  \gamma c \hskip 3cm
d\gamma = \gamma ( d-hc) $$\\
$$ a \delta = ( \delta + h\gamma ) ( a - hc ) \hskip 1cm
b\delta = (\delta - h \gamma) (b - hd) +h (\delta + h \gamma) (a-hc)) $$
$$ c \delta =  ( \delta + h \gamma ) c \hskip 3cm
d\delta = ( \delta -h \gamma ) d + h ( \delta + h\gamma ) c $$\\
$$ a \beta = (\beta + h\alpha)( a + hc ) -
h ( \delta + h \gamma) (a-hc) $$
$$ b \beta =  ( \beta - h\alpha) ( b + hd ) - h ( \delta - h\gamma) ( b-
 hd) + $$
$$ h \Bigl( ( \beta + h\alpha) ( a + hc ) - h
 ( \delta + h\gamma) ( a- hc)v\Bigr )  $$
$$c\beta = \Bigl ( ( \beta + h \alpha ) + h ( \delta + h\gamma )\Bigr )c $$
$$d\beta = \Bigl ( ( \beta - h \alpha ) + h ( \delta - h\gamma )\Bigr ) (d+
hc) $$\\
Either by taking differentials of both sides of the above equations or by
solving (25), the relations
between the one forms are obtained as:
$$ \alpha^2 = h \gamma \alpha \hskip 1cm \beta^2 = h\Bigl ( ( \alpha + \d
elta)\beta
+ h \alpha \delta\Bigr )  $$
$$ \gamma ^2 = 0 \hskip 3cm \delta ^2 = h \gamma \delta $$\\
$$ \alpha \beta + \beta \alpha = - h ( \beta\gamma + \alpha \delta )
\hskip 2cm
\alpha \gamma  + \gamma \alpha = 0 \hskip 2cm  \delta \alpha + \alpha \d
elta  = h ( \alpha + \delta ) \gamma $$
$$ \gamma\delta + \delta\gamma = 0 \hskip 2cm \beta \delta + \delta \beta =
h ( \alpha \delta + \gamma \beta )  $$
$$ \gamma\beta  + \beta \gamma = h \gamma ( \alpha + \delta )  \hskip 2cm
\alpha \delta + \delta \alpha = h ( \alpha + \delta  ) \gamma $$

{\large \bf  The Cartan Maurer Forms in
$\Omega [ SL_h(2) ] $}

The structure of defining relations of $ \Omega [ SL_h (2) ] $ take a
 much simpler form if one changes
the basis to the Cartan-Maurer forms. By introducing $ \Omega = S(T)dT $
it follows from (21)
and (25) that
 \begin{equation} \Omega_1 T_2 R_{21} = T_2 R_{21}\Omega \end{equation}
\begin{equation} \Omega_1 R_{12}\Omega_2 R_{21}  + R_{12} \Omega_2 R_
{21} \Omega_1 = 0  \end{equation}
We have also
$$ d \Omega + \Omega \Omega = 0 $$
It is convenient to take $ \Omega $ as
$ \Omega  = \left( \begin{array}{ll} { \omega_+ + \omega_-
 \over 2 } & u \\ v  \over 2 } & u \\ v
& { \omega_+ - \omega_- \over 2 }   \end{array}\right) $
The result of solutions of (26) and (27)
are :
{}From (26) we have:
$$ [ \omega_+ , a ]  = 0 \hskip 2cm [ \omega_+ , b ] = 4h^2 av  $$
$$ [ \omega_+ , c ]  = 0 \hskip 2cm [ \omega_+ , d ] = 4h^2 cv $$\\
$$ [ \omega_- , a ]  = 2hav \hskip 2cm [ \omega_- , b ]  = -2hbv $$
$$ [ \omega_- , c ]  = 2hcv \hskip 2cm [ \omega_- , d ]  = -2hdv $$\\
$$ [ v , a ]  = 0 \hskip 2cm [ v , b ]  = 2hav $$
$$ [ v , c ]  = 0 \hskip 2cm [ v , d ]  = 2hcv $$\\
$$ [ u , a ]  = -ha (\omega_- + hv ) \hskip 2cm
[ u , b ]  = -2ha (u-h^2v ) + hb( \omega_- - hv ) $$
$$ [ u , c ]  = -hc (\omega_- + hv ) \hskip 2cm  [ u , d ]  = -2hc (u-h^2v)
+ hd ( \omega_- - hv ) $$
{}From (27) we have:
$$ v^2 = \omega_{\pm}^2 = \{ \omega_+ , \omega_- \} = \{ \omega_{\pm} . v
\} =
0 $$
$$ u^2 =  h \Bigl ( (u+hv)(\omega_+ - \omega_- ) + ( \omega_+ +
\omega_- ) ( u
+ hv ) \Bigr ) $$\\ $$ \{ \omega_+ , u \} = -4h^2 \omega_- v $$
$$ \{ \omega_- , u \} = 2h  ( uv - vu ) $$
        \vskip 1cm

{\large \bf Appendix A :  Derivatives in  Non Commutative
Differential Calculus }

In this appendix we show that a covariant diffrential calculus on the
quantum
planes can be constructed only if the derivatives act from the right .
Since our discussion applies also to $ SL_q(2)$ and the
Wess-Zumino
differencial calculus on the associated quantum plane $ R_q(2) $ , we
discuss this case in detail which is generaly more familiar.
For $ SL_q(2) $ the defining relations are:
$$ ab = q ba \hskip 2cm cd = q dc $$
 \begin{equation} ac = q ca \hskip 2cm bd = q db \end{equation}
$$ bc= cb \hskip 2cm ad - da = ( q-q^{-1}) bc $$
with the quantum determinant $ D = ad - qbc = da - q^{-1} bc = 1 $ and the
associated quantum plane relations
 \begin{equation} xy = q yx  \ \ \ , \xi^2 = \eta^2 = \xi\  \eta +
\ q^{-1} \ \eta \ \xi \ = 0\  \end{equation}
Consider the change of coordinates
 \begin{equation} x' = a x + b y \hskip 2cm y' = c x + d y \end{equation}
Interpreting the symbols $ \partial_x $ and $ \partial_y$  as derivatives
acting from the left and demanding the validity of the chain rule we have:
 \begin{equation} \partial_x = a \partial_ {x'} + c \partial_{y'} \end
{equation}
 \begin{equation}\partial_y = b \partial_ {x'} + d \partial_{y'}
\end{equation}
Formal inversion of (30) yields:
 \begin{equation} x = d x' - q^{-1} b y' \hskip 2cm y = -q c x' + a y' \
end{equation}
Again chain rule gives:
 \begin{equation} \partial_x' = d \partial_ {x} - q c \partial_{y} \en
d{equation}
 \begin{equation} \partial_y' = -q^{-1} b \partial_ {x} + a \partial_{
y}\end{equation}
Now two problems arise:

1) Relations (34,35) are not the inverses of (31,32) . In fact inserti
on of (34,35) into
the right hand side of (31) yields:
 \begin{equation} R.H.S.\ \  of \ \ (31) = ( ad - q^{-1} cb) \partial_x
+ (ca - qac) \partial_y \ne
\partial_x  \end{equation}
2) The obvious relations $ \partial_x (x) = \partial_y ( y ) = 1 $ and $
 \partial _x (y) = \partial _y ( x ) = 0 $ are not covariant ;
indeed we have:
 \begin{equation} \partial_{x'} ( x') = ( d\partial_x - q c \parti
al_y ) ( a x + b y ) = da -
qcb \ne 1\end{equation}   Both of the above difficulties disappear  if w
e act the
derivatives from the right and change the relations correspondingly:
 \begin{equation} \partial_x =  \partial_ {x'} a +  \partial_{y'} c \
end{equation}
 \begin{equation} \partial_y =  \partial_ {x'} b +  \partial_{y'} d \e
nd{equation}
and
 \begin{equation} \partial_x' = \partial_ {x} d - q  \partial_{y} c \end
{equation}
 \begin{equation} \partial_y' = -q^{-1} \partial_ {x} b +  \partial_{y} a\
end{equation}
The analogues of (36) and (37) now become respectively :
 \begin{equation} R.H.S.\ \  of \ \ (38) = \partial ( da - q^{-1} bc)
+ \partial_y (-q ca + ac)
\partial_y =
\partial_x  \end{equation}
and
 \begin{equation} ( x')\partial_{x'}  =  ( ax + by ) ( \partial_x d -
q  \partial_y c)
=  ad - qbc = 1 \end{equation}
Other relations of this kind can be checked similarly. In the general
 case this
asymetry between left and right differentiation stems from the fact
that for a
quantum matrix $ T $ the relation $ (T^{-1})^t =  (T^{t})^{-1} $ where
$ T^t$
is the transpose of $ T $, is not valid .
\vspace {1cm}

ACKNOWLEDGEMENT : The author wishes to express his deep gratitude to
F. Ardalan for inspiration ,
encouragement and the final reading of the manuscript.
\newpage

REFERENCES:
\begin{enumerate}
\item See Takhtajan L. A. and references therein; Introduction to quantum
groups and integrable
massive models of quantum field theory ,M.L.Ge and B. H. Zhao,eds.
World Scientific (1991)
\item Zakrzewski,S. ; Lett. Math. Phys. {\bf {22}} 287-289 (1991)
\item Faddeev, L. D.,Reshetikhin, N. Yu. and Takhtajan, L. A.; Leningrad
Math. J {\bf {1}} 193-225 (1990)
\item Ohn, C. H.; Lett. Math. Phys. {\bf {25}} 85-88 (1992)
\item Manin, Yu. Bonn preprints MPI/91-47; MPI/91-60 (1991)
\item Wess,J. and Zumino, B. Nuclear Phys. B (proc. Supp.) {\bf {18} B},
302 (1990)
\item Woronowics S.L.Commun. Math. Phys. {\bf {122}},125 (1989)
\item Aref'eva ,I.Ya.; and Volovich I. V.; Phys. Lett.{\bf{B
291}},273-278, (1992)
\item Schwenk J., and Wess, J. ;  Phys. Lett.{\bf{B 268}},179-188, (1991)
\item Schupp,P.;Watts,P. and Zumino, B. Lett. Math. Phys. {\bf {25}}
139-147 (1992)
\end{enumerate}
\end{document}